\documentclass[twocolumn,aps,pra,longbibliography,floatfix,superscriptaddress,preprintnumbers,10pt]{revtex4-1}
\usepackage{color}
\usepackage{textcomp}
\usepackage{amsmath}
\usepackage{amssymb}
\usepackage{graphicx}
\usepackage{esint}
\usepackage[normalem]{ulem}

\makeatletter
\@ifundefined{textcolor}{}
{%
 \definecolor{BLACK}{gray}{0}
 \definecolor{WHITE}{gray}{1}
 \definecolor{RED}{rgb}{1,0,0}
 \definecolor{GREEN}{rgb}{0,1,0}
 \definecolor{BLUE}{rgb}{0,0,1}
 \definecolor{CYAN}{cmyk}{1,0,0,0}
 \definecolor{MAGENTA}{cmyk}{0,1,0,0}
 \definecolor{YELLOW}{cmyk}{0,0,1,0}
}

\usepackage{times}
\usepackage{hyperref}
\hypersetup{colorlinks=true}
\usepackage[all]{hypcap} 

\newcommand{\dLdag}{d^{\dagger}_{L}}
\newcommand{\dL}{d^{\vphantom{\dagger}}_{L}}

\newcommand{\llangle}{\left\langle}
\newcommand{\rrangle}{\right\rangle}

\begin{document}

\title{Stabilization of a Majorana Zero Mode through Quantum Frustration}

\author{Gu Zhang}
\affiliation{Department of Physics, Duke University, P.O. Box 90305, Durham, North Carolina 27708, USA}
\affiliation{Institute for Quantum Materials and Technologies, Karlsruhe Institute of Technology, 76021 Karlsruhe, Germany}
\email{gu.zhang@kit.edu}

\author{Harold U. Baranger}
\email{baranger@phy.duke.edu}

\affiliation{Department of Physics, Duke University, P.O. Box 90305, Durham, North Carolina 27708, USA}

\begin{abstract}
We analyze a system in which a topological Majorana zero mode (tMZM) combines with a Majorana produced by quantum frustration (fMZM) to produce a novel ground state. 
The system that we study combines two parts, a grounded topological superconducting wire that hosts two tMZMs at its ends,  and an on-resonant quantum dot connected to two dissipative leads. The quantum dot with dissipative leads creates an effective two-channel Kondo (2CK) state in which quantum frustration yields an isolated fMZM at the dot. We find that coupling the dot to one end of the topological wire stabilizes the tMZM at the other end.
Three routes are used to obtain these results: (i)~calculation of the conductance through an auxiliary detector quantum dot, (ii)~renormalization group (RG) arguments and the g-theorem, and (iii)~a fully non-equilibrium calculation of the $I(V)$ curve and shot noise $S(V)$ through the detector dot. 
In addition to providing a route to achieving an unpaired Majorana zero mode, this scheme provides a clear signature of the presence of the 2CK frustration-induced Majorana.
\end{abstract}

\date{May 28, 2020}
\maketitle

%
%

%


\section{Introduction}

Electron states with topological character 
and quantum frustration from competing interactions 
are two major themes in current condensed matter physics. In both contexts, fractionalized degrees of freedom on the boundary of a system occur. Perhaps the best known example is the possibility of Majorana zero modes (MZMs) at the ends of a one-dimensional (1D) system. MZMs are exotic self-conjugate edge states that can occur through either topology  
\cite{DasSarmaMajoranaQInf15,AguadoReview17,LutchynNatRevMat18,StanescuTopoBook} or fine-tuning of competing interactions \cite{AffleckTCK92,EmeryKivelsonPRB92,WongAffleck94,Affleck2IKPRB95,[{}][{, Chapter 28.}]GogolinBook}. 
Here we study the interplay between a topological Majorana zero mode (tMZM) and frustration-induced Majorana (fMZM) in a nanoscale system of quantum dots and wires. We show that a fMZM can stabilize a tMZM. 

Quantum frustration typically produces states of matter that are delicately balanced between competing options and which then show fractionalization \cite{SchifferRMP13}. 
Quantum impurity models---an interacting quantum system coupled to leads---provide several canonical examples. (Note that 
quantum impurity models are effectively 1D since the impurity couples to a limited set of states in the leads \cite{WilsonRMP75}.) The two channel Kondo (2CK) model, for instance, has been extensively studied \cite{GogolinBook,NozieresBlandin80,ZawadowskiPRL80,EmeryKivelsonPRB92,AffleckTCK92}: 
an impurity spin is equally coupled to two metallic leads. It would be natural for the impurity spin to form a singlet with each lead, but it cannot because of entanglement exhaustion. This frustration in screening the impurity leads to a non-Fermi-liquid ground state in which there is a degeneracy of $\sqrt{2}$ at the impurity. This signals fractionalization and the existence of an unpaired fMZM \cite{MZM_pair}. It has also been discussed in the two impurity Kondo model  \cite{JayaprakashPRL81,Affleck2IKPRB95} and the dissipative resonant level model \cite{Mebrahtu13,Zheng1-GPRB14}. Experimentally, several groups have investigated in detail nanoscale systems with an unpaired fMZM of this type \cite{Potok2CK07,Mebrahtu12,Mebrahtu13,KellerDGGNat15,IftikharPierreNat15,IftikharPierreScience18}--- to date, the fine tuning required appears to be easier to achieve than the creation of a topological state.

Topological MZMs, in addition to their inherent interest, have attracted attention because their non-Abelian statistics provide a possible route toward fault tolerant quantum computation \cite{DasSarmaMajoranaQInf15,PachosBook12}. To construct and observe such MZMs, researchers have proposed multiple systems 
\cite{LutchynNatRevMat18,SatoAndo-TopSupRPP17,SunJia-MajVortexNQM17,HaimOreg-TopSupPRep19,Motome-HuntingJPSP20}.
One particularly promising 1D system consists of a semiconducting nanowire made of a material that has strong spin-orbit coupling which is placed in proximity to a \textit{s}-wave superconductor and in a magnetic field \cite{LutchynNatRevMat18}. In this system, signatures of tMZM through measurement of the conductance have been intensively pursued \cite{LutchynNatRevMat18}. 

In contrast to the free elementary particle predicted by Majorana, the effective tMZMs in condensed matter always appear in pairs in finite size systems \cite{DasSarmaMajoranaQInf15,LutchynNatRevMat18}. Unfortunately, tMZMs lose many of their interesting properties when they hybridize with their partners. 
The inter-MZM coupling decays as $\propto \exp(-L/\xi)$ \cite{DasSarmaMajoranaQInf15,LutchynNatRevMat18}, with $L$ the distance and $\xi$ the superconducting correlation length in the nanowire. Experimentally, 
this hybridization can not generally be ignored \cite{LutchynNatRevMat18}.
Consequently, to see the full effect of a tMZM, a method to stabilize the tMZM against inter-MZM coupling is desirable.

In this paper, we stabilize a tMZM against hybridization with its partner by coupling that partner to an unpaired Majorana fermion of a dissipative quantum dot. 
Through this stabilization, the frustration-induced fMZM of the $R \!=\! R_Q  \!\equiv\!  h/e^2$ dissipative resonant level model can be experimentally detected. 

We emphasize that ``stabilization" here is understood in the renormalization group (RG) sense. In the absence of stabilization, we show that a finite inter-tMZM coupling, no matter how small, is RG \emph{relevant}: it effectively increases with decreasing temperature and thus drastically changes the system ground state. 
In contrast, when the dissipation-induced fMZM is present, the inter-tMZM  coupling is \emph{irrelevant} and vanishes at zero temperature, regardless of its bare value.
Because of this RG aspect, the stabilization studied here is qualitatively different from other proposals where the bare inter-tMZM coupling can be
reduced to zero through fine tuning.
See Refs.\,\cite{AguadoReview17, deng_majorana_2016, ElsaPRB17, ClarkePRB17, PradaNpjQM17} for examples of fine tuning by increasing the nanowire size, coupling the nanowire to a quantum dot, or through electronic interactions.

\begin{figure}[t]
\includegraphics[width=6.3cm]{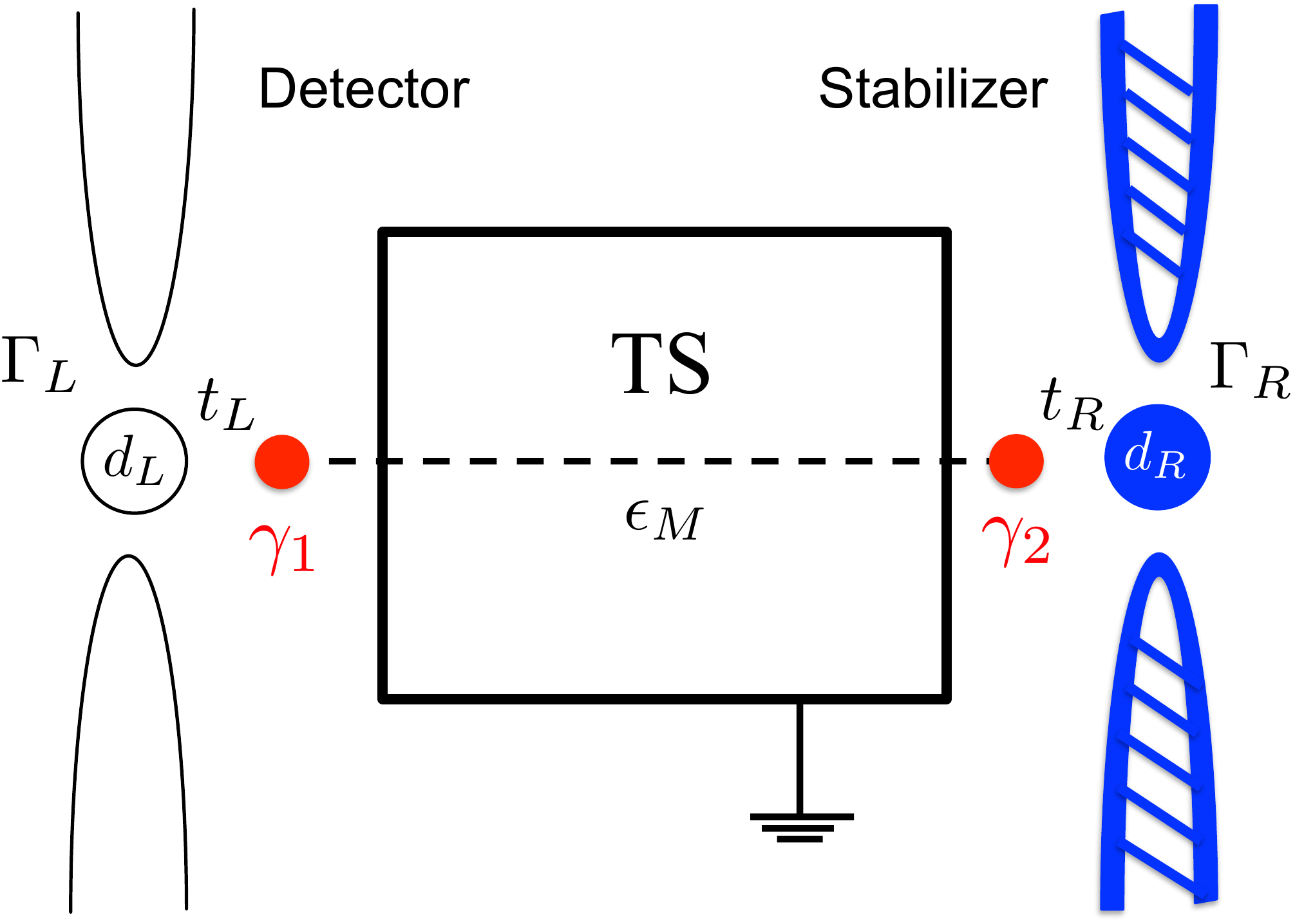}
\caption{The structure of the system. Two tMZMs $\gamma_1$ and $\gamma_2$ are realized at two ends of a nanowire on top of a grounded topological superconductor (TS). We calculate the conductance through the left quantum dot to detect the existence of the tMZM $\gamma_1$. The right quantum dot, through which transport is dissipative (blue), couples to $\gamma_2$ and thereby stabilizes $\gamma_1$ as an isolated tMZM.
}
\label{fig:structure}
\end{figure}

The rest of the paper is organized as follows. We begin with the introduction of our model in 
Sec.\,\ref{sec:model}.
After that,  Sec.\,\ref{sec:review} contains a brief review of the dissipative resonant level model and its frustration-induced  Majorana fermion.
With those ingredients, we calculate in Sec.\,\ref{sec:conductance} the conductance through the detector quantum dot (the left side of Fig.\,\ref{fig:structure}) that couples to one of the tMZM. These results allow us to conclude that the fMZM of a dissipative resonant level model stabilizes the tMZM by coupling to its partner., thus leaving a single unpaired tMZM. 
We further interpret the result with the g-theorem of boundary conformal field theory in Sec.\,\ref{sec:g_theorem}, where we stress the importance of the dissipation-induced fMZM in the process of  stabilization. 
To gain further information and support for our view, we find the non-equilibrium conductance and shot noise through the detector quantum dot with full counting statistics methods in Sec.\,\ref{sec:full_counting}.
Our results indicate that there is an abrupt transition of the ground state when the fMZM joins the system. This transition can be clearly observed through conductance, shot noise, as well as the Fano factor of the detecting resonant level.
Finally, we summarize our paper in Sec.\,\ref{sec:summary}.

\section{The System}
\label{sec:model}

The system we consider consists of three major parts: (i) a superconducting nanowire that hosts two tMZMs, (ii) a resonant level that detects the presence of a tMZM by its conductance, and (iii) a dissipative resonant level, formed in a quantum dot, which introduces a fMZM that stabilizes the signature of the tMZM.

We consider the superconducting nanowire as a bare bones system, shown in Fig.\,{\ref{fig:structure}}, that has a pair of tMZM, $\gamma_1$ and $\gamma_2$ (red dots), at its ends. The coupling between these two tMZMs is $\epsilon_M \neq\! 0$ and hence the Hamiltonian of the system is 
\begin{equation}
H_\textrm{sys.} = i \epsilon_M \gamma_1 \gamma_2 .
\label{eq:Hsys}
\end{equation}
The goal is to have effectively $\epsilon_M\!=\!0$ so that even at zero temperature the topological feature of $\gamma_1$ is evident.

In order to assess directly whether $\gamma_1$ is indeed an independent tMZM, we incorporate a detector explicitly. As the presence of a tMZM affects many physical properties, different types of detectors could be used. We choose to consider the conductance through a spinless quantum dot modeled as a resonant level \cite{DongPRBR11}, pictured on the left of Fig.\,{\ref{fig:structure}}. As there are no interactions here, the Hamiltonian of the detector is simply
\begin{equation}
\begin{aligned}
H_\textrm{detect.} = &\; \epsilon_L^{\vphantom{\dagger}} d^\dagger_L d^{\vphantom{\dagger}}_L + 
\sum_{k,\alpha} \epsilon_k c^{\dagger}_{k L \alpha} c^{\vphantom{\dagger}}_{k L \alpha} \\
& + V_L \sum_{k, \alpha} \left( c^{\dagger}_{k L \alpha} d^{\;}_L + d^\dagger_L c^{\vphantom{\dagger}}_{k L \alpha} \right) ,
\end{aligned}
\label{eq:Hdetect}
\end{equation}
where $\epsilon_{L}^{\vphantom{\dagger}}$ is the dot energy level, $c_{k L \alpha}^{\vphantom{\dagger}}$ is for electrons in the $\alpha=$ $S$ (source) or $D$ (drain) lead, and $V_L$ is the dot-lead coupling. We assume that the dot is tuned to resonance, $\epsilon_L \!= 0$, and symmetrically coupled to the leads. 

\begin{figure}[t]
\includegraphics[width=7.2cm]{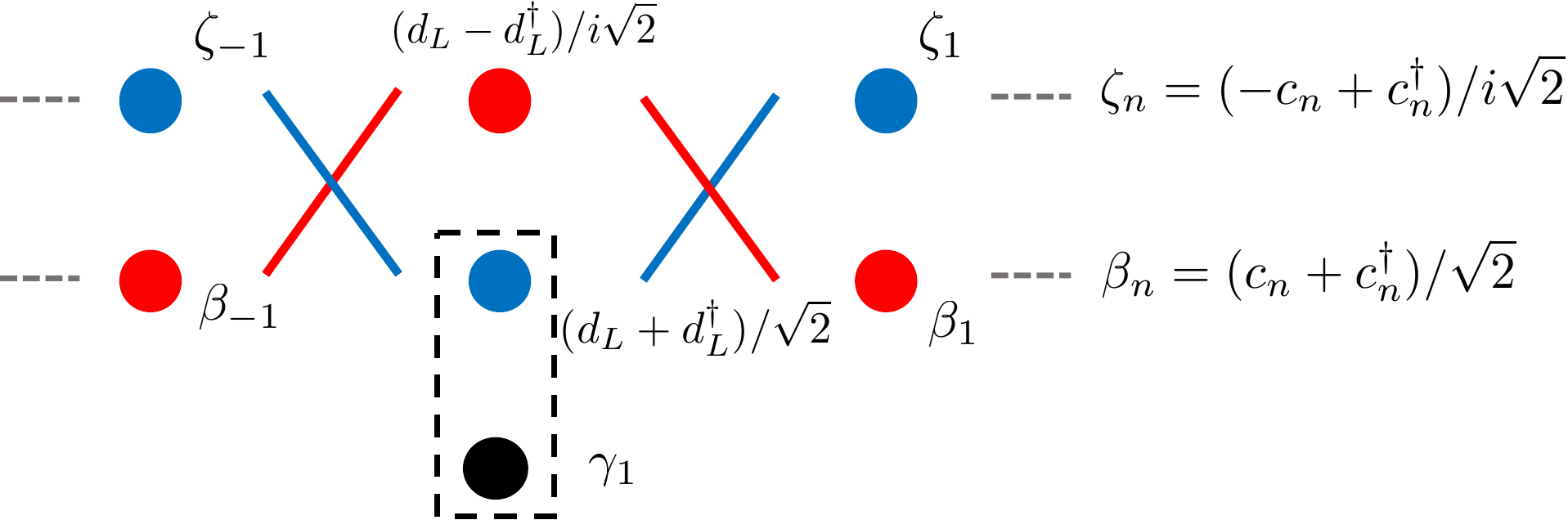}
\caption{A resonant level (with operators $c_n$ and $c_n^{\dagger}$ for the $n$th site in the leads) couples to a tMZM $\gamma_1$.
When fine-tuning, the resonant level system can be considered as two independent Majorana chains \cite{DongPRBR11} (indicated by red and blue colors, respectively).
The tMZM $\gamma_1$ and $(d_L + d_L^{\dagger})/\sqrt{2}$ form into a singlet (indicated by the black dashed box), thus breaking one of the Majorana channels (the blue one).
The conductance becomes $e^2/2h$, totally from the red Majorana channel.
}
\label{fig:lead_majoranas}
\end{figure}

The tMZM $\gamma_1$ is tunnel coupled to a combination of $d_L$ and $d^\dagger_L$. Because $\epsilon_L = 0$, all such combinations are equivalent and so we take
\begin{equation}
H_\textrm{sys.-det.} = i t_L \frac{d^{\dagger}_L + d_L}{\sqrt{2}} \gamma_1 .
\label{eq:sys_det}
\end{equation}
The (zero-temperature) conductance through the left dot, denoted $G_L$, for $H=H_\textrm{sys.}+H_\textrm{detect.}+H_\textrm{sys.-det.}$ has been studied previously \cite{DongPRBR11}. 
There is a clear signature of the coupling to the topological Majorana: while $G_L = e^2/h$ on resonance for a classic resonant level ($t_L=0$), in the presence of the tMZM one obtains half that value, $G_L \!=\! e^2/2h$. Intuitively, the tMZM hybridizes with part of the resonant level and so blocks the conductance of a ``half chain'', as illustrated in Fig.\,\ref{fig:lead_majoranas} \cite{DongPRBR11}. 
Thus, we shall use $G_L \!=\! e^2/2h$ as the sign that a tMZM is present. 

However, for any non-zero $\epsilon_M$, the conductance reverts to the topologically trivial $G_L \!=\! e^2/h$ at low temperature, $T\ll\epsilon_M$. 
Indeed, $\epsilon_M$ is RG relevant, as shown below, and so grows large at low temperature regardless of its bare value (i.e., the value determined by the experimental system). The resulting extreme sensitivity to the value of $\epsilon_M$ makes observation of the tMZM especially difficult. 


In order to stabilize $\gamma_1$, one natural idea is to couple its partner $\gamma_2$ to an isolated Majorana fermion, thereby removing it from potentially hybridizing with $\gamma_1$. 
Fortunately, such an isolated Majorana fermion is known to exist in the 2CK quantum impurity model. Specifically, the frustration between the two channels leaves an effective fMZM $(d+d^{\dagger})/\sqrt{2}$ untouched, where $d \equiv i S_x - S_y$ is the effective fermionic operator and $S_{\sigma}$ are the Pauli matrices of the impurity spin  \cite{EmeryKivelsonPRB92,ShillerPRB95,GogolinBook}.
However, this effective fMZM is fundamentally a \emph{spin} operator and so will not naturally couple with the \emph{spatial} degree of freedom $\gamma_2$. 

To solve this problem, we propose to generate an isolated fMZM with a dissipative quantum dot \cite{Mebrahtu13,Mebrahtu12}. The corresponding resonant level model with ohmic impedance $R \!=\! R_Q$ \cite{Mebrahtu13,Zheng1-GPRB14} is known to be equivalent to the 2CK model as well as the Luttinger liquid resonant level model with Luttinger liquid interaction $g = 1/2$. The quantum dot forming the resonant level hosts a real \cite{real-mf} decoupled Majorana fermion \cite{Mebrahtu13,Zheng1-GPRB14}. Since this fMZM involves a \emph{spatial} degree of freedom, we can model its coupling to $\gamma_2$ with the standard inter-Majorana coupling.

\section{Quantum Frustration from Dissipation: The Dissipative Resonant Level Model}
\label{sec:review}

We begin by sketching the needed elements of the theory of the dissipative resonant level model \cite{Mebrahtu12,Mebrahtu13,Zheng1-GPRB14,LiuRLdissipPRB14}, emphasizing the formation of its dissipation-induced fMZM.
The dissipative resonant level model describes a quantum dot that couples to two dissipative spin-polarized leads. It is defined by the Hamiltonian \cite{IngoldNazarov92,NazarovBlanterBook,LiuRLdissipPRB14}
\begin{equation}
\begin{aligned}
H_\textrm{dissip.}= & H_{\text{dot}} + H_{\text{lead}} + H_{\text{T}} \\
=&\; \epsilon_R d^\dagger d^{\vphantom{\dagger}} + 
\sum_{k,\alpha} \epsilon_k c^{\dagger}_{k \alpha} c^{\vphantom{\dagger}}_{k \alpha} 
\hfill \\
& + \sum_{k, \alpha}  V_{\alpha} \left( e^{-i\phi_\alpha} c^{\dagger}_{k \alpha} d^{\;} + e^{+i\phi_\alpha} d^\dagger c^{\vphantom{\dagger}}_{k \alpha} \right),
\end{aligned}
\label{eq:Hdissip}
\end{equation}
where $\alpha \!=\! S, D$ for the source and drain, respectively.
The notation parallels that for the detector: $\epsilon_R$ is the dot energy level, $c^\dagger_{k\alpha}$ creates an electron in the lead labeled $\alpha$, and lead $\alpha$ couples to the dot with strength $V_{\alpha}$. The key aspect of the model is the coupling to dissipation in the tunneling term [third line of Eq.\,\eqref{eq:Hdissip}]. 

In modeling the dissipation, we follow a standard approach \cite{IngoldNazarov92,NazarovBlanterBook,VoolDevoretIJCTA17}. The phase fluctuation operator $\phi_\alpha$ is conjugate to the charge fluctuation operator for the capacitor between the dot and lead $\alpha$. Thus, the operator $e^{\pm i\phi_\alpha}$ in Eq.\,(\ref{eq:Hdissip}) accounts for the change in charge upon tunneling. The current and voltage fluctuations caused by the electrons tunneling on and off the dot excite the ohmic environment of the leads. This environment is modeled as a bath of harmonic oscillators \cite{CaldeiraLeggett81,IngoldNazarov92,NazarovBlanterBook,VoolDevoretIJCTA17}. Because it is the charge moving across the dot that excites the environment, the difference $\varphi\equiv \phi_S - \phi_D$ couples to the harmonic oscillators. The bath causes the correlation function of these fluctuations to be
\begin{equation}
\langle e^{-i\varphi(t)}e^{i\varphi(0)}\rangle \propto (1/t)^{2r}
\label{eq:dissipative_correlations}
\end{equation}
where the exponent $r$ is related to the resistance of the environment by 
$r \equiv Re^2/h \equiv R/R_Q$. With this correlation function, the conductance and scaling behavior can be found. 


Interactions are thus introduced by the dissipation [i.e.\ the third line of Eq.\,(\ref{eq:Hdissip}) is \emph{not} quadratic]---theoretically, we are faced with an \emph{interacting quantum impurity} model. It is then natural to proceed with a bosonization treatment followed by renormalization (RG)
\cite{GogolinBook,GiamarchiBook,Anderson70,Cardy81,KaneFisherPRB92,EggertAffleck92,Mebrahtu13,Zheng1-GPRB14}.


Following the standard technique \cite{GogolinBook,LiuRLdissipPRB14}, we unfold each lead into an infinite chiral fermion channel and  then apply 
chiral bosonization,
\begin{equation}
c_{\alpha} (x) = \frac{F_{\alpha}}{\sqrt{2\pi a}} e^{i\phi_{\alpha} (x)+ik_Fx},
\label{eq:bosonization}
\end{equation}
where $F_{\alpha}$ is the Klein factor $\left\{ F_{\alpha}, F_{\alpha'} \right\} = 2 \delta_{\alpha,\alpha'}$ that preserves the fermionic commutation relations, and $a$ is the lattice constant. 
The bosonic field operator $\phi_{\alpha}$, representing the collective modes of the corresponding chiral channel \cite{GogolinBook}, has the standard commutation relation
\begin{equation}
[\partial_x \phi_{\alpha}(x), \phi_{\alpha'}(x') ] = i\delta_{\alpha,\alpha'} \pi \delta (x - x').
\label{eq:commutators}
\end{equation}
We further define fields in the common ($\phi_c$) and difference ($\phi_f$) sectors through the rotation
\begin{equation}
\phi_f = \frac{\phi_{ S} - \phi_{D}}{\sqrt{2}},\ \ \ \ \phi_c = \frac{\phi_{S} + \phi_{D}}{\sqrt{2}}.
\end{equation}
The fields $\phi_c$ and $\phi_f$ reflect the dot occupation number and the electron number difference between two leads, respectively.

Substituting these bosonization expressions into the Hamiltonian (\ref{eq:Hdissip}), one finds that the bosonic field $\phi_f$ and dissipative phase $\varphi$ appear in the same way in the tunneling term. As they both have power-law correlation functions [Eq.\,\eqref{eq:dissipative_correlations} and \cite{GogolinBook}], we therefore combine them with the transformation 
\begin{equation}
\begin{aligned}
\phi_f' & \equiv \frac{1}{\sqrt{1+r}} \left( \phi_f + \frac{1}{\sqrt{2}} \varphi \right)  \\
\varphi' & \equiv \frac{1}{\sqrt{1+r}} \left( \sqrt{r} \phi_f + \frac{1}{\sqrt{2 r}} \varphi \right),
\end{aligned}
\label{eq:boson_rotation}
\end{equation}
through which the 
tunneling Hamiltonian becomes \cite{Zheng1-GPRB14,LiuRLdissipPRB14}
\begin{equation}
\begin{aligned}
H_{\text{T}} & = \sum_{\alpha \in \{S,D\}} \frac{V_{\alpha}}{\sqrt{2 \pi a}} \left( e^{- i \frac{1}{\sqrt{2}} \phi_c} e^{- i\alpha \sqrt{\frac{1+r}{2}} \phi_f'} F_{\alpha} d + \text{h.c.} \right),
\end{aligned} 
\label{eq:ht_effective}
\end{equation}
where $\alpha \!=\! \pm 1$ for source and drain, respectively.

Eq.\,\eqref{eq:ht_effective} effectively mimics the tunneling Hamiltonian of a Luttinger liquid  
in which the interaction in the common ($c$) and difference ($f$) sectors is different. 
(For related work on links between the physics of Luttinger liquids and dissipative tunneling see, e.g., Refs.\,\cite{MatveevGlazman93,FlensbergPRB93,SassettiWeissEPL94,SafiSaleurPRL04,LeHurLiPRB05,JezouinPierre13}.)
Following the well-established RG technique for Luttinger liquids \cite{GogolinBook,KaneFisherPRB92,GiamarchiBook}, we obtain RG equations when $\epsilon_R \!=\! 0$ \cite{KaneFisherPRB92,LiuRLdissipPRB14}: 
\begin{equation}
\begin{aligned}
\frac{dV_S}{d\ln\tau_c} & = \left[1 - \Big( \frac{1+r}{4} + \frac{K_1}{4} +\frac{K_2}{2} \Big) \right] V_S, \\
\frac{dV_D}{d\ln\tau_c} & = \left[1 - \Big( \frac{1+r}{4} + \frac{K_1}{4} -\frac{K_2}{2} \Big) \right] V_D, \\
\frac{dK_1}{d\ln\tau_c} & = - 4\tau_c^2 \Big[ \left(V_S^2 + V_D^2 \right) K_1 + \left(V_S^2 - V_D^2 \right) K_2  \Big], \\
\frac{dK_2}{d\ln\tau_c} & = - 2\tau_c^2 \Big[ \left(V_S^2 + V_D^2 \right) K_2 + \left(V_S^2 - V_D^2 \right) \Big],
\end{aligned}
\label{eq:rg_equations}
\end{equation}
where $\tau_c$ is the energy cutoff that decreases gradually with the decreasing temperature, and $K_1$, $K_2$ are the fugacity parameters that incorporate the symmetric and anti-symmetric parts of the dot-$\phi_c$ interaction during the RG flow.
Initially, $K_1 \!=\! 1$ and $K_2 \!=\! 0$.
Importantly, for finite asymmetry $V_S \!-\! V_D$, $|K_2|$ increases, which in turn leads to increased asymmetry upon RG flow. Generically, the flow thus ends at a ground state in which the quantum dot is completely hybridized with either the source or the drain (and cut from the other), depending on whether $V_S$ or $V_D$ is initially larger. The transition between these two candidate ground states, known as a boundary quantum phase transition, has been experimentally realized \cite{Mebrahtu12,Mebrahtu13}.

Non-trivial behavior appears at the quantum critical point $V_S \!=\! V_D$: \emph{frustration} between hybridization with the source versus the drain prevents the quantum dot from being fully hybridized.
In fact, a finite residual entropy $\ln\!\sqrt{1+r}$ remains at zero temperature \cite{WongAffleck94}.
This residua; entropy mimics that of the 2CK problem at the intermediate fixed point \cite{EggertAffleck92,Mebrahtu13,Zheng1-GPRB14,Crossover}, where a spin Majorana becomes isolated due to overscreening.
Specially, when $R\!=\!R_Q$ (i.e.\ $r\!=\!1$), the residual entropy at the fine-tuned quantum critical point becomes $\ln\!\sqrt{2}$, which coincides with that of a Majorana fermion.
\emph{This Majorana is the fMZM we use in this paper to stabilize the tMZM.}

For $R\!=\!R_Q$ the model has been thoroughly investigated through bosonization and refermionization  \cite{Mebrahtu13,Zheng1-GPRB14,Crossover}, following that for the 2CK model 
\cite{EmeryKivelsonPRB92,SenguptaGeorges94,ShillerPRB95}
or a $g \!=\! 1/2$ Luttinger liquid resonant level model \cite{GogolinBook}.
Following their example, we apply the unitary transformation
\begin{equation}
    U = e^{i(d^{\dagger} d - \frac{1}{2}) \phi_c(0)/\sqrt{2}},
    \label{eq:u-trans}
\end{equation}
to remove the common field $\phi_c$ from the tunneling term.
However, this unitary transformation introduces two minor side effects.
First, the impurity operator is now dressed with the common field $\phi_c(0)$, 
\begin{equation}
    d \to d e^{i K_1 \frac{1}{\sqrt{2}}\phi_c(0)},
\end{equation}
with $K_1 \!=\! 1$ initially, i.e.\ before the RG flow of Eq.\,\eqref{eq:rg_equations}.
Second, the unitary transformation introduces a quartic interaction 
\cite{Zheng1-GPRB14,LiuRLdissipPRB14},
\begin{equation}
    H_{\text{extra}} = -\frac{v}{2\sqrt{2}} \left(d^{\dagger}d - \frac{1}{2}\right) \partial_x \phi_c(x)\Big|_{x = 0}
    \label{eq:quartet}
\end{equation}
where $v$ is the Fermi velocity, that couples $\phi_c$ to the impurity occupation number.

Strictly, the phase factor $\exp[i\phi_c(0)/\sqrt{2}]$ that attaches to $d$ as well as the interaction Eq.\,(\ref{eq:quartet}) are quite important at high temperatures \cite{Zheng1-GPRB14}.
However, $K_1$ decreases according to the RG equations \eqref{eq:rg_equations} \cite{KaneFisherPRB92}, so that at low temperature $d\exp[iK_1\phi_c(0)/\sqrt{2}]$ and the bare operator $d$ become indistinguishable. With regard to the induced density-density interaction (\ref{eq:quartet}), it has scaling dimension 3/2 (see, for instance, \cite{SchillerHershToulousePRB98} or \cite{GanPRB95} where similar terms have been encountered) and is thus RG irrelevant. Consequently, as we are only interested in the low temperature physics near the ground state, \emph{both} the phase attached to impurity operators and the quartic interaction can be safely neglected.

Finally, we define a Majorana representation for the degree of freedom represented by  $d$: $\chi_1 \equiv ( d^{\dagger} \!+\! d)/\sqrt{2}$ and $\chi_2 \equiv i(d^{\dagger} \!-\! d )/\sqrt{2}$. Because of the unitary transformation mentioned in the last paragraph, this is no longer simply the dot level but rather a nonlinear mixture of the dot and the density in the two leads near the dot. Both resulting MZMs are highly localized near the quantum dot.

For the specific case $r\!=\!1$, the dependence on the difference field $\phi_f'$ in \eqref{eq:ht_effective} can be expressed as a fermionic operator $\psi_f  \!\equiv\!  e^{i\phi_f'}/\sqrt{2\pi a}$ (using the Klein factor from the original bosonization). 
The result of these manipulations is an effective Majorana Hamiltonian for the right-hand dot and leads:  
\begin{equation}
\begin{aligned}
H_\textrm{dissip.} = & \sum_{k} \epsilon_k \psi^{\dagger}_{f,k}\psi_{f,k} + (V_S - V_D) \frac{\psi^{\dagger}_f(0) - \psi_f(0)}{\sqrt{2}} \chi_1 \\
&  + i(V_S + V_D) \frac{\psi^{\dagger}_f (0) + \psi_f(0)}{\sqrt{2}} \chi_2 + i \epsilon_R \chi_1 \chi_2, 
\end{aligned}
\label{eq:Majorana}
\end{equation}
Straightforwardly, at the quantum critical point where $\epsilon_R = 0$ and $V_S \!=\! V_D$, one impurity Majorana $\gamma_1$ becomes isolated, thus leading to the $\ln\! \sqrt{2}$ residual entropy.

In the rest of the paper, we focus on the symmetric point $V_S = V_D \equiv V_R$, and couple this system to the right end of the superconducting nanowire as a stabilizer. 


\section{Conductance in the Detector: Three Cases}
\label{sec:conductance}

With the system introduced, we calculate the conductance $G_L$ through the left quantum dot (the detector) in different scenarios. 

\subsection{No Stabilizer}

For the simplest scenario, without the presence of any stabilizer, $\gamma_2$ couples only to its partner $\gamma_1$. This case has been studied previously \cite{DongPRBR11}: the non-trivial zero-temperature conductance $e^2/2h$ abruptly becomes the trivial one $e^2/h$ upon \emph{any} non-zero $\epsilon_M$.
From the RG perspective, this means that $\epsilon_M$ is relevant and thus when temperature decreases  $\epsilon_M$ effectively \emph{increases}.

\subsection{Frustration-Induced Degeneracy in Right Dot}
\label{sec:Cond-dissip}

With the presence of the dissipative resonant level, the key final ingredient in our problem is the connection between the right dot and the topological wire. This is simply tunneling, as for the left dot Eq.\,(\ref{eq:sys_det}); for detailed discussions of tunneling between tMZM and those arising from Klein factors in bosonization see, e.g., Refs.\,\cite{Beri-MajoranaKleinPRL13,HerviouPRB16,GiulianoAffleckNPB19}.
Generically, $\gamma_2$ couples to both $\chi_1$ and $\chi_2$, yielding the Hamiltonian
\begin{equation}
\begin{aligned}
 H_\textrm{sys.-dis.} = 
 + i t_{R1} \,\gamma_2 \,\chi_1 + i t_{R2} \,\gamma_2 \,\chi_2 
\end{aligned}
\label{eq:sys_dis}
\end{equation}
with arbitrary couplings $t_{R1}$ and $t_{R2}$.



The full Hamiltonian for our problem,
$H \!=\! H_\textrm{sys.} \!+\! H_\textrm{detect.} \!+\! H_\textrm{dissip.} \!+\! H_\textrm{sys.-det.} \!+\! H_\textrm{sys.-dis.}$,
is quadratic and so can be solved through the equation of motion method. We calculate the 
conductance of the left quantum dot that probes the $\gamma_1$ tMZM. With symmetric coupling, its equilibrium conductance is related to the dot spectral function by 
\begin{equation}
G_L = -\Gamma_L \frac{e^2}{h} \int\frac{d\omega}{2 \pi}  \text{Im}  \left\{ G^R(\dL,\dLdag) (\omega) \right\} \partial_{\omega}n_F(\omega),
\label{eq:conductance_spectrum}
\end{equation}
where $G^R(\dL,\dLdag)(\omega)$ is the Fourier transform of the retarded Green function 
$-i \theta(t) \big\langle \big\{ d_L(0), d_L^{\dagger}(t) \big\} \big\rangle$, $n_F(\omega)$ is the Fermi distribution function, and $\Gamma_L \!=\! \pi \rho_0 V_L^2$ is the level broadening.

The retarded Green function of the left dot from the equation of motion method \cite{Bruus-Flensberg} is
\begin{equation}
G^R(d_L,d^{\dagger}_L) (\omega) = \frac{1}{\omega + i\Gamma_L - \epsilon_L - \Sigma (\omega)},
\label{eq:igf_dissipativeless}
\end{equation}
where the self-energy $\Sigma (\omega)$ incorporates the effects of the coupling between 
(i) the left dot and $\gamma_1$, 
(ii) $\gamma_1$ and $\gamma_2$ Eq.\,\eqref{eq:Hsys}, and 
(iii) $\gamma_2$ and the right dot Eq.\,\eqref{eq:sys_dis}). 
With the presence of the dissipative stabilizer, the self energy is 
\begin{equation}
\begin{aligned}
&\Sigma (\omega)^{-1} = \frac{\omega}{t^2_L} - \frac{1}{\omega + \epsilon_L + i\Gamma_L} \\
& - \frac{\epsilon_M^2}{t^2_L} \left(\omega - \frac{t^2_{R1}}{\omega+ \epsilon_R + i\eta} - \frac{t^2_{R1}}{\omega- \epsilon_R + i\eta} \right)^{-1}\\
&- \frac{\epsilon_M^2}{t^2_L} \left(\omega - \frac{t^2_{R2}}{\omega+ \epsilon_R + i\Gamma_R} - \frac{t^2_{R2}}{\omega- \epsilon_R + i\Gamma_R} \right)^{-1},
\end{aligned}
\label{eq:dissipative_self_energy}
\end{equation}
where $\eta $ is a positive infinitesimal and $\Gamma_R  \!=\! \pi\rho_0 V_R^2$ is the level broadening of the right dot in the absence of dissipation.

With Eqs.\,\eqref{eq:igf_dissipativeless} and \eqref{eq:dissipative_self_energy}, the conductance through the left dot when $\epsilon_R =0$ is
\begin{equation}
G_L = \frac{1}{2} \frac{e^2}{h},
\label{eq:conductance_with_right_system}
\end{equation}
\emph{independent} of the values of any parameters (such as $t_{R2}$ or $\epsilon_M$). This striking independence implies, for instance, that fine tuning of the coupling between $\gamma_2$ and the right dot is \emph{not} needed, a significant experimental simplification. This result holds only within the validity of our model, of course: one should have $\epsilon_M \!\ll\! \Delta$ in order to have the tMZM pair ($\Delta$ is the superconducting gap in the proximitized nanowire) and $\epsilon_M \!\ll\!\Gamma_R$ in order to have a fMZM from frustration. 

The conductance Eq.\,(\ref{eq:conductance_with_right_system}) is one of our main results. It indicates that the introduction of the $R \!=\! R_Q$ dissipative quantum dot stabilizes $\gamma_1$. 

We emphasize that the stabilization of $\gamma_1$ refers to the fact that at zero temperature $\epsilon_M$ always effectively vanishes. This is true even if $\epsilon_M$ is significant initially, such as in a system with a short nanowire. This complete stabilization is uniquely guaranteed by the presence of the fMZM: it couples with $\gamma_2$ into a singlet, thus preventing the inter-tMZM coupling.
We further illustrate this point in Sec.\,\ref{sec:g_theorem} below through analysis with the g-theorem.


Fine-tuning of the energy level of the quantum dot is required only for the right-hand dot, $\epsilon_R \!=\! 0$. We do not need a fine-tuned left dot since $\epsilon_L$ is irrelevant at the nontrivial fixed point:
at this point, $\gamma_1$ and $(d^{\dagger}_L  \!+\! d_L)/\sqrt{2}$ form a singlet, thus strongly suppressing the hybridization between  $(d^{\dagger}_L  \!+\! d_L)/\sqrt{2}$ and $(d^{\dagger}_L  \!-\! d_L)/i\sqrt{2}$.
Experimentally, the irrelevance of $\epsilon_L$ is a signature of the tMZM: instead of the usual Lorentzian shape from the resonant level model, the zero temperature conductance of the 
left-hand dot 
is expected to be \emph{flat} as a function of $\epsilon_L$. For non-zero temperature, the conductance will be constant as long as $\epsilon_L(T) < \Gamma_L(T)$, both of which may vary with temperature because of renormalization effects. 

\subsection{Dissipation-Free 
Right Dot}
\label{sec:dissipation_free}

To highlight the role of dissipation, we now consider what happens if there is no dissipation in the right-hand leads, $r \!=\!0$. Thus the full-transmission fixed point Hamiltonian Eq.\,(\ref{eq:Majorana}) is replaced by a second copy of the resonant level Hamiltonian Eq.\,(\ref{eq:Hdetect}).
Since the Hamiltonian remains quadratic, we again use the equation of motion method to find the retarded Green function of the left dot; the explicit form of the self-energy now becomes
\begin{equation}
\begin{aligned}
&\Sigma (\omega)^{-1}  = \frac{\omega}{t^2_L} - \frac{1}{\omega + \epsilon_L + i\Gamma_L} \\
&- \frac{\epsilon_M^2}{t^2_L} \left(\omega - \frac{t^2_R}{\omega+ \epsilon_R + i\Gamma_R} - \frac{t^2_R}{\omega- \epsilon_R + i\Gamma_R} \right)^{-1}.
\end{aligned}
\label{eq:impurity_self_interaction}
\end{equation}
In the absence of dissipation, the two MZMs on the right dot become equivalent, allowing us to freely choose $\gamma_2$ to couple to any linear combination of them with the coupling strength $t_R$. The remaining MZM will be hybridized by the leads.


The conductance of the left quantum dot in this case is 
\begin{equation}
G_L = \frac{e^2}{h} \frac{2 t_L^2 t_R^2 + \epsilon_M^2 \Gamma_L \Gamma_R}{4 t^2_L t^2_R + \epsilon^2_M \Gamma_L \Gamma_R},
\label{eq:conductance_dissipativeless}
\end{equation}
where $\Gamma_R \!=\! \pi \rho_0 V_R^2$ is the broadening of the right resonant level.
Eq.\,(\ref{eq:conductance_dissipativeless}) displays an interesting feature: 
the equilibrium zero temperature conductance varies \emph{continuously} between the trivial ($e^2/h$) and the non-trivial ($e^2/2h$) values, depending on the details of the system.
This crossover originates from the competition between the dot-MZM couplings and the hybridization of the quantum dots by the leads.

Eq.\,\eqref{eq:conductance_dissipativeless} implies that $\epsilon_M$ is effectively controllable through fine-tuning $t_{L,R}$ and $\Gamma_{L,R}$. Indeed, this result agrees with previous investigations of quantum dots coupled to proximitized nanowires in which non-local effects produced by the two tMZM 
(which are related directly to the value of $\epsilon_M$) are tunable through fine-tuning of the dot level \cite{AguadoReview17, deng_majorana_2016, ElsaPRB17, ClarkePRB17}. 
We stress that in these systems the effective inter-tMZM overlap is only reduced.
In strong contrast, with the fMZM induced by dissipation in our system, this overlap 
is driven by interactions to \emph{zero}. 

\subsection{Explanation with g-Theorem}
\label{sec:g_theorem}

To summarize briefly thus far, we have shown that the conductance through the MZM-coupled left quantum dot is strongly influenced by the nature of the system on the right. (i) In the absence of a right-hand system, the two MZMs $\gamma_1$ and $\gamma_2$ couple into a trivial state for which $G_L/(e^2/h) = 1$. (ii) When the right-hand system is present but without dissipation, the conductance through the left dot is between the trivial and nontrivial values, depending on the details of the system. (iii) Finally, the nontrivial state is stabilized when the right-hand system is dissipative with $R \!=\! R_Q$, leading to the zero-temperature conductance $G_L/(e^2/h) = 1/2$.

Through the g-theorem, we now provide a simple way to understand these results. As the counterpart of the famous c-theorem of two-dimensional conformal field theory \cite{ZomolodchikovJETPLett83, CardyPLB88}, the g-theorem treats boundary phase transitions and relates the stability of the fixed points to the impurity or boundary entropy. Specifically, if the bulk parameters remain invariant during the RG flows, the flow will bring the system toward the fixed point with a smaller impurity entropy \cite{AffleckGTheoremPRL91,FriedanKonechnyPRL04}. We have calculated the ground state degeneracy of our system at the two fixed points in the three scenarios above. The results are compiled in Table\,\ref{tab:summary}, and we now discuss each scenario in turn.

\begin{table}[t]
    \begin{tabular}{| c | c | c | c |}
    \hline
     & $t_{R} = 0$ & $t_R \neq 0$ and $R = 0$ & $t_R\neq 0$ and $R = R_Q$ \\ \hline
    $g_{\text{trivial}}$ & 1 & 1 & $\sqrt{2}$ \\ \hline
    $g_{\text{nontrivial}}$ & $\sqrt{2}$ & 1 & 1 \\ \hline
     $G_L/(e^2/h)$ & 1 & $\dfrac{2 t_L^2 t_R^2 + \epsilon_M^2 \Gamma_L \Gamma_R}{4 t^2_L t^2_R + \epsilon^2_M \Gamma_L \Gamma_R}$ & 1/2 \\ \hline
    \end{tabular}
    \caption{System characteristics at different fixed points. Here, 
    $g_{\text{trivial}}$ and $g_{\text{nontrivial}}$ are the degeneracies of the trivial and nontrivial fixed points, respectively, and $G_L$ is the zero temperature conductance through the left dot, 
For simplicity we have used $t_R \!=\! 0$ to label the decoupling of the right dot.}
     \label{tab:summary}
\end{table}

(i) If the right-hand system is absent ($t_R \!=\! 0 $) and $\epsilon_M \!\neq 0$, the trivial fixed point is non-degenerate: the two tMZMs $\gamma_1$ and $\gamma_2$ form into a singlet and the left quantum dot is completely hybridized with the leads. In contrast, the nontrivial fixed point has a decoupled tMZM, namely $\gamma_2$, yielding a ground state degeneracy $\sqrt{2}$. The g-theorem then implies, in agreement with the conductance calculation above, that the nontrivial fixed point is unstable. 
Alternatively, we notice that the leading operator at the nontrivial fixed point is the hybridization between the leads and 
$(\dLdag \!+\! \dL)/\sqrt{2}$, which has the scaling dimension $1/2$. Consequently, the hybridization operator is relevant and sabotages the nontrivial fixed point. 

(ii) If the right-hand system is a dissipationless quantum dot, the ground states of both the trivial and nontrivial fixed points are non-degenerate.
Consequently, the operator that connects these two fixed points is marginal, leading to a crossover between the trivial and nontrivial fixed points. This crossover is reflected in the intermediate value of the conductance, Eq.\,(\ref{eq:conductance_dissipativeless}).
From an RG point of view, because the parity of the superconducting island is conserved, tunneling happens simultaneously in the left and right quantum dots (see Appendix for details). Thus the scaling dimension of the hybridization doubles compared to case (i), rendering it marginal.

(iii) Finally, when the right quantum dot is dissipative, at the nontrivial fixed point the isolated fMZM  $\chi_2$ couples to $\gamma_2$ in a singlet, thus leading to a non-degenerate ground state.
In contrast, at the trivial fixed point, $\chi_2$ remains isolated, yielding degeneracy $\sqrt{2}$.
Consequently, the g-theorem predicts that the RG flow brings the system toward the \emph{nontrivial} fixed point. 
Alternatively, the lead-dot hybridization is suppressed by the dissipation and so has a larger scaling dimension than in case (ii). The hybridization thus becomes irrelevant (for any $R \neq 0$) and the Majorana feature is protected.
This protection uniquely exists in a dissipative or an interacting system where a non-trivial (i.e., the Majorana-like) residual entropy has been added to the system through the frustration between two competing dissipative leads.

\section{Full Counting Statistics}
\label{sec:full_counting}

In the previous section, we investigate the zero-temperature linear-response conductance through the left quantum dot. Our calculation indicates that the effective $\epsilon_M$ exactly vanishes when $\gamma_2$ couples to a dissipative quantum dot, 
thus completely stabilizing $\gamma_1$ from the inter-tMZM coupling.
In this section, we instead investigate the \textit{non-equilibrium} current and shot noise of the model at different fixed points with full counting statistics \cite{LevitovReznikovPRB04,GogolinKomnikPRB06, KamenevBook}.
Since analysis in previous sections indicates that $t_{R2}$ is RG irrelevant in the presence of finite $t_{R1}$, we take $t_{R2} \!=\! 0$ for simplicity.

\subsection{Full Counting Statistic in the Majorana-Fermion-Coupled Resonant Level Model}

In full counting statistics, the current and noise are calculated through \cite{LevitovReznikovPRB04}
\begin{equation}
I = \frac{e \llangle \delta q \rrangle}{\tau},\ \ \ \text{and} \ \ \ 
S = \frac{2e^2 \llangle\delta^2 q \rrangle}{\tau},
\label{eq:current_and_noise_from_fcs}
\end{equation}
where the moments 
\begin{equation}
\llangle \delta^n q \rrangle = (-i)^n  \frac{\partial^n}{\partial \lambda^n} \ln \chi(\lambda) \Big\vert_{\lambda = 0}
\label{eq:fcs_presciption}
\end{equation}
characterize the charge correlations. 
In Eq.\,\eqref{eq:fcs_presciption},
the generating function $\chi(\lambda) = \sum_{q} e^{i q \lambda} P_q (\tau)$ includes tunneling events to all orders ($q\in Z^{\ge}$ is a non-negative integer), where $\lambda$ is the measuring field and $P_q(\tau)$ is the probability that charge $q e$ tunnels through the barrier(s) during the measuring time $\tau$. Practically, in 1D systems $\ln \chi(\lambda) \!=\! -i \tau U(\lambda,-\lambda)$ where $U(\lambda,-\lambda)$ is the adiabatic potential.

For the left quantum dot of our system (the detector), the adiabatic potential of the resonant level model is
\begin{equation}
\begin{aligned}
\frac{\partial}{\partial_{\lambda_-}} U(\lambda_-,\lambda_+) &\\ 
 = \frac{V^2_L}{2} \int d\omega & \left[e^{-i (\lambda_- - \lambda_+)/2} G^{< }(d_L, d_L^{\dagger}) g_{L\alpha}^{+-} \right. \\
 & \left. - e^{i (\lambda_- - \lambda_+)/2} g_{L\alpha}^{- + } G^{>}(d_L, d_L^{\dagger}) \right],
\end{aligned}
\label{eq:adiabatic_potential}
\end{equation}
where $G^{>}(d, d^{\dagger})$ [$G^{<}(d, d^{\dagger})$] is the full greater (lesser) impurity Green function and $g_{L\alpha}$ is the bare Green function for the source or drain lead ($\alpha \!=\! S$ or $D$). The  four components of the bare lead Green function are given by \cite{GogolinKomnikPRB06,KamenevBook}
\begin{equation}
\begin{aligned}
g^{--}_{L\alpha} (\omega) & = g_{\alpha}^{++} (\omega) = i 2 \pi \rho_0 (n_{\alpha} -1/2), \\
g_{L\alpha}^{-+} (\omega) & = i 2\pi \rho_0 n_{\alpha}, \\
g^{+-}_{L\alpha}(\omega) & = -i 2\pi \rho_0 ( 1 - n_{\alpha}),
\end{aligned}
\label{eq:bare_lead_gfs}
\end{equation}
where $n_{S,D} = n_F(\epsilon \pm V/2)$ is the distribution function of the leads and $V>0$ is the bias applied between source and drain.
At zero temperature, these distribution functions simplify to $n_S = \Theta(-\epsilon - V/2)$ and $n_D = \Theta(-\epsilon + V/2)$.

To calculate the full Green functions $G^{<}$ and $G^{>}$, we divide the Hamiltonian into two parts 
$H = H_0 + \delta H $, 
where $H_0 = H_{\text{sys.}} + H_{\text{detect.}} + H_{\text{dissip.}} + H_{\text{sys.-dis.}}$ contains the non-equilibrium resonant level model (the detector) plus the other parts of the system, 
while $\delta H = H_{\text{sys.-det}}$ connects these two parts.

When $\delta H\!=\!0$ ($t_L \!=\! 0$), these two parts are disconnected, and we can calculate the Green functions of them separately.
The non-equilibrium Green function matrix $G_0(d_L,d^{\dagger}_L)$ of the resonant level Hamiltonian $H_\textrm{detect.}$ is known \cite{GogolinKomnikPRB06},
\begin{equation}
\begin{aligned}
G_0(d_L,d_L^{\dagger})(\omega) & =\left[\begin{array}{cc}
  G^R_0 (d_L,d_L^{\dagger})(\omega) &  G^<_0 (d_L,d_L^{\dagger})(\omega) \\
  G^>_0 (d_L,d_L^{\dagger})(\omega) & G^A_0 (d_L,d_L^{\dagger})(\omega)
  \end{array}\right] \\
&=\frac{1}{\omega^2 + \Gamma_L^2 e^{i\lambda}}\Big(                 
  \begin{array}{cc}
  \omega - i \Gamma_L &  i e^{i \lambda} \Gamma_L \\
  -i \Gamma_L & -\omega - i \Gamma_L,
  \end{array}
\Big), 
\end{aligned}
\label{eq:free_gf}
\end{equation}
with the four entries referring to the retarded, lesser, greater and advanced Green functions, respectively (from top to bottom, left to right). Note that the measuring field $\lambda$ appears. 
The remaining parts of the system ($H_{\text{sys.}} \!+\! H_{\text{dissip.}} \!+\! H_{\text{sys.-dis.}}$), on the other hand, remain in equilibrium when $t_L \!=\! 0$. We thus obtain the equilibrium retarded Green function of $\gamma_1$, 
\begin{equation}
G_0^R(\gamma_1,\gamma_1)(\omega) = \frac{\omega^2 - t_R^2}{\omega (\omega^2 - t_R^2 - \epsilon_M^2)}.
\label{eq:rest_part_free_gf}
\end{equation}

With $\delta H$, we calculate the full lesser Green function through the Dyson equation \cite{KamenevBook,GogolinKomnikPRB06,SelaNoneqQdotsPRB09}
\begin{equation}
\begin{aligned}
G^< & = G^R \Sigma^< G^A \\
&+ (1 + G^R \Sigma^R) G^<_0 (1 + \Sigma^A G^A).
\end{aligned}
\label{eq:gf_retarded_relation}
\end{equation}
After including all possible processes, Eq.\,\eqref{eq:gf_retarded_relation} becomes 
\begin{widetext}
\begin{equation}
\begin{aligned}
G^<(d_L,d_L^{\dagger})(\omega) & = G^<_0(d_L,d_L^{\dagger})(\omega) + G^R(d_L,\gamma_1) \Sigma^R_{\gamma_1,d_L} G_0^<(d_L,d_L^{\dagger}) \Sigma^A_{d_L^{\dagger},\gamma_1} G^A(\gamma_1,d_L^{\dagger})(\omega) \\
&\ \ \ \ + G^R(d_L,\gamma_1) \Sigma^R_{\gamma_1,d^{\dagger}_L} G_0^<(d^{\dagger}_L,d_L) \Sigma^A_{d_L,\gamma_1} G^A(\gamma_1,d_L^{\dagger})(\omega) \\
&\ \ \ \ + G^R(d_L,\gamma_1)(\omega) \Sigma^R_{\gamma_1,d_L} G^<_0 (d_L,d_L^{\dagger}) + G^<_0(d_L ,d_L^{\dagger}) \Sigma^A_{d_L^{\dagger}, \gamma_1} G^A(\gamma_1,d_L^{\dagger})(\omega),
\end{aligned} 
\label{eq:second_relation}
\end{equation}
\end{widetext}
where the interaction terms $\Sigma^A_{d_L^{\dagger},\gamma_1} \!=\! \Sigma^R_{\gamma_1,d_L}  \!=\! -\Sigma^R_{\gamma_1,d_L^{\dagger}} \!=\! - \Sigma^A_{d_L,\gamma_1}\!=\!  t_L$, and the impurity function $G_0^<(d_L^{\dagger},d_L)$ equals
\begin{equation}
\begin{aligned}
G_0^<(d_L^{\dagger},d_L)(\omega) & = \frac{-i \Gamma_L e^{-i\lambda}}{\omega^2 +\Gamma_L^2 e^{-i\lambda}}.
\end{aligned}
\label{eq:several_free_gfs}
\end{equation}
Eq.\,(\ref{eq:second_relation}) contains two additional Green functions, 
$G^R(d_L,\gamma_1)(\omega)$ and $G^A(\gamma_1,d_L^{\dagger})(\omega)$,
that can also be found from Dyson equations:
\begin{equation}
\begin{aligned}
G^R(d_L,\gamma_1) (\omega) & = 0 + G^R_0(d_L,d_L^{\dagger})(\omega) \Sigma_{d_L^{\dagger},\gamma_1} G^R(\gamma_1,\gamma_1)(\omega) , \\
G^R(\gamma_1,\gamma_1)(\omega) & = G^R_0(\gamma_1,\gamma_1)(\omega) \\
&+ G^R_0(\gamma_1,\gamma_1)(\omega) \Sigma_{\gamma_1,d_L} G^R(d_L,\gamma_1)(\omega) \\
&+ G^R_0(\gamma_1,\gamma_1)(\omega) \Sigma_{\gamma_1,d_L^{\dagger}} G^R(d_L^{\dagger},\gamma_1)(\omega) , \\
G^R(d_L^{\dagger},\gamma_1)(\omega) & = 0 + G^R_0(d_L^{\dagger}, d_L)(\omega) \Sigma_{d_L,\gamma_1} G^R(\gamma_1,\gamma_1)(\omega).
\end{aligned} 
\label{eq:linear_equations}
\end{equation}

Eqs.\,\eqref{eq:free_gf}-\eqref{eq:linear_equations} yield the lesser Green function required in Eq.\,\eqref{eq:adiabatic_potential}. 
Following analogous steps, we also obtain the greater Green function. Combining these with the 
bare lead Green functions Eq.\,\eqref{eq:bare_lead_gfs},
we obtain the adiabatic potential and the generating function from Eq.\,\eqref{eq:adiabatic_potential}. This finally allows evaluation of the the current and noise, Eq.\,\eqref{eq:current_and_noise_from_fcs} \cite{NoteBook}.

Below we present expressions for the non-equilibrium current and noise in different cases corresponding to different fixed points. Though from the calculation sketched here, 
we have the full nonlinear dependence on the bias $V$, for simplicity we expand the expressions to leading order in $V$ \cite{NoteBook}, following the assumption that the bias $V$ is smaller than all other relevant energy scales.

\subsection{Non-Interacting tMZMs ($\epsilon_M = 0$)}
\label{sec:non_interacting_M}

We begin with the simple case $\epsilon_M \!=\! 0$.
In this case, $\gamma_1$ is stable since it totally decouples from its partner $\gamma_2$, and the result thus is independent of the presence of a stabilizer.
The current calculated from the adiabatic potential in this case is
\begin{equation}
I(V ) = \frac{e^2}{h} \left[\frac{V}{2} - \left(\frac{1}{\Gamma_L^2} - \frac{\Gamma_L^2}{4t_L^4} \right) \frac{e^2}{24} V^3 + \mathcal{O}(V^4) \right],
\label{eq:current_decoupled_MFs}
\end{equation}
where $\mathcal{O}(V^4)$ indicates the neglected higher-order terms. Note that in agreement with Eq.\,(\ref{eq:conductance_dissipativeless}) the linear-response conductance from this calculation is $e^2/2h$. As discussed above and in \cite{DongPRBR11}, this comes about because one of the left dot's Majorana degrees of freedom is fully coupled to the tMZM while the other fully hybridizes with the lead to form a transparent half-channel. 


Interestingly, the cubic term [$\mathcal{O}(V^3)$] in (\ref{eq:current_decoupled_MFs}) displays a  competition between two processes: (i) backscattering in the transparent half-channel  which reduces the current and (ii) hybridization of the left leads with $(d_L \!+ d^{\dagger}_L)/\sqrt{2}$, pulling it away from the tMZM and thereby enhancing the current.
More specifically, when $\Gamma^2 \!>\! 2 t_L^2$ the hybridization is stronger so that the $\mathcal{O}(V^3)$ current increases with an increasing bias and vice versa.
Note that the presence of this competition requires two Majorana channels in opposite limits---one completely healed and the other totally disconnected. To the best of our knowledge, such a situation exists only in systems that are topological.   


Turning to the fluctuations of the current, we find that the zero-temperature shot noise to leading order is 
\begin{equation}
S = 2\frac{e^3}{h} \left[\frac{V}{4} + \left(-\frac{1}{\Gamma_L^2} - \frac{\Gamma_L^2}{4 t_L^4} + 
\frac{1}{t_L^2} \right)\frac{e^2}{48} V^3 + \mathcal{O}(V^4) \right],
\label{eq:shot_noise_decoupled}
\end{equation}
with the leading term proportional to bias.
This linear term is a signal that the transmission is \emph{not} perfect: for a system with perfect transmission, the leading term should instead be $\propto V^3$ [see Eq.\,\eqref{eq:shot_noise_interacting_MFs} for an example when $\epsilon_M \neq 0$].

The competing effects seen in the nonlinear current also appear here.
Indeed, noise to the next-leading order [i.e., $O(V^3)$] reaches it maximum when two competing processes equal (i.e., $\Gamma^2 \!=\! 2t_L^2$).
This point, with half transmission probability, is known to have the largest shot noise \cite{IhnBook}.

Near equilibrium, the Fano factor, $F\!\equiv\! S/2eI$, becomes $1/2$, implying that the current is carried by quasi-particles with effective charge $e^* \!=\! e/2$ at zero temperature.
As in the charge 2CK case discussed in \cite{LandauCornfeldSelaPRL18}, here the $e^* \!=\! e/2$ fractional charge property originates from the fact that one of the Majorana channels in the leads completely decouples (Fig.\,\ref{fig:lead_majoranas}).

\subsection{Interacting tMZMs without Stabilization}
\label{sec:interacting_M_decoupled_R}

Now we add back the interaction between two tMZMs ($\epsilon_M \!\neq\! 0$), but do not include the 
right-hand quantum dot 
($t_R = 0$) as the stabilizer.

The nonlinear current calculated with full counting statistics now becomes
\begin{equation}
I (V) = \frac{e^2}{h} \left[V - \frac{1 + 2 t_L^2/\epsilon_M^2 + 2t_L^4/\epsilon_M^4}{12\Gamma_L^2}e^2 V^3 + \mathcal{O}(V^4) \right],
\label{eq:current_interacting_MFs}
\end{equation}
with the trivial equilibrium conductance $e^2/h$, in agreement with Ref.\,\cite{DongPRBR11}.
In contrast to Eq.\,(\ref{eq:current_decoupled_MFs}), there is no sign of competing effects in the nonlinear term. Indeed, at a perfectly conducting fixed point, non-equilibrium effects necessarily reduce the conductance, so the sign of the $V^3$ term is fixed. 

The shot noise now becomes
\begin{equation}
S =2 \frac{e^3}{h} \left[ \left(\frac{ t_L^4}{\epsilon_M^4} + 1 \right) \frac{e^2}{12 \Gamma_L^2} V^3 + \mathcal{O}(V^4) \right],
\label{eq:shot_noise_interacting_MFs}
\end{equation}
with the leading term $\propto\! V^3$.
Eq.\,(\ref{eq:shot_noise_interacting_MFs}) contains both back\-scattering-induced noise and noise generated by the coupling between $\gamma_1$ and $i(-d_L \!+\! d^{\dagger}_L)$.
Notice the factor $\epsilon_M^4$ in the denominator here as well as in the nonlinear current Eq.\,\eqref{eq:current_interacting_MFs}. The high power indicates that $\epsilon_M$ is RG relevant and drives the system to a different fixed point when it is non-zero.

Since the linear term in the shot noise vanishes, here the Fano factor can be defined as the ratio between shot noise and twice the value of the ``backscattering current" as in the charge 2CK model \cite{LandauCornfeldSelaPRL18}. Using the $V^3$ terms in both the current and the shot noise, Eqs.\,(\ref{eq:current_interacting_MFs}) and (\ref{eq:shot_noise_interacting_MFs}), we see that the Fano factor depends on the ratio $t_L /\epsilon_M$. We prefer not to discuss an effective charge of the quasi-particle in this case since the inelastic contributions to the nonlinear terms complicate the interpretation \cite{SelaNoiseKondoPRL06}. 
The variation of the Fano factor indicates that at the quantum dot, an electron either backscatters or tunnels into the superconducting island. 


\subsection{Interacting tMZMs Stabilized by Frustration}

Comparison of the results in these two cases (Sections \ref{sec:non_interacting_M} and \ref{sec:interacting_M_decoupled_R}) supports our analysis that the coupling $\epsilon_M$ between the two  tMZMs is relevant, leading to the destruction of the Majorana signature in the detector. In this section we add the frustration-induced Majorana fermion $\chi_1$ and show that it stabilizes the probed tMZM $\gamma_1$.

In this case, the model in Section \ref{sec:Cond-dissip} yields the current
\begin{equation}
\begin{aligned}
I (V) = \frac{e^2}{h} &\left[\frac{V}{2} +  \left(\frac{\epsilon_M^4 \Gamma_L^2}{t_L^4 t_R^4} + \frac{2\epsilon_M^2 \Gamma_L^2}{t_L^4 t_R^2} + \frac{\Gamma_L^2}{t_L^4} - \frac{4}{\Gamma_L^2} \right) \frac{e^2}{96} V^3 \right. \\  &  \qquad + \mathcal{O}(V^4) \Big],
\end{aligned}
\label{eq:current_tr_on}
\end{equation}
with the expected linear-response conductance $e^2/2h$, Eq.\,\eqref{eq:conductance_with_right_system}. Note further that the full $I(V)$ reduces to Eq.\,(\ref{eq:current_decoupled_MFs}) when $\epsilon_M \!=\! 0$.
The fact that $\epsilon_M$ is in the numerator in the nonlinear term shows that it is RG irrelevant. This is, then, verification of our analysis that the presence of $\chi_1$ stabilizes $\gamma_1$ against the inter-tMZM coupling $\epsilon_M$.

The shot noise now becomes
\begin{equation}
\begin{aligned}
S & = 2 \frac{e^2}{h} \left[ \frac{V}{4} + \left( - \frac{1}{4 \Gamma^2} - \frac{\Gamma^2}{16 t_L^4} + \frac{1}{4t_L^2}  \right. \right. \\
& \left.\left. -\frac{\epsilon_M^4 \Gamma^2}{16 t_L^4 t_R^4} - \frac{\epsilon_M^2 \Gamma^2}{8 t_L^4 t_R^2} + \frac{\epsilon_M^2}{4 t_L^2 t_R^2}  \right) \frac{e^2}{12} V^3 + \mathcal{O}(V^4) \right]. 
\end{aligned}
\label{eq:shot_noise_tr_on}
\end{equation} 
In the $V^3$ term, the inter-tMZM coupling $\epsilon_M$ again appears only in the numerator, showing its RG irrelevance. This allows a smooth $\epsilon_M \to 0$ limit yielding, indeed, 
Eq.\,(\ref{eq:shot_noise_decoupled}). The shot noise here does have a fixed linear term. As in the $\epsilon_M \!=\! 0$ case, from the Fano factor we deduce tunneling of quasi-particles with effective charge $e^* \!=\! e/2$.  

These results for both the nonlinear transport $I(V)$ and the shot noise, then, support our equilibrium result (Section \ref{sec:conductance}) that $\chi_1$ stabilizes the tMZM $\gamma_1$ against its coupling to $\gamma_2$.

\section{Summary}
\label{sec:summary}

The effect that we elucidate here arises in essence from combining a Majorana degree of freedom due to topology with a Majorana produced by fine-tuned quantum frustration (i.e.\ interactions). This is a highly unusual situation: we connect a topological system with one that is not topological and find no unusual feature at the boundary (and in particular no MZM).


We arrive at this result by, first, calculating the conductance through a detector quantum dot and, second, supporting the calculation through RG arguments and the g-theorem.
The combination of the two types of MZM originates from the preference of the system for a non-degenerate ground state, arrived at by coupling the topological Majorana $\gamma_2$ and the quantum dot Majorana fermion $\chi_1$ into a singlet. 
Our conclusions are further supported by a fully non-equilibrium calculation in which the nonlinear current and shot noise reveal signatures of the RG relevant and irrelevant processes. 


Perhaps the most interesting implication of our results for future work is that 
the long-ignored frustration generated Majorana fermion is in principle detectable. 
Indeed, it could be potentially beneficial for quantum computation.


\bigskip
\emph{Acknowledgements---}
We thank Ruixing Zhang for fruitful discussions.
The work at Duke was supported by the U.S.\ DOE Office of Science, Division of Materials Sciences and Engineering, under Grant No.\,{DE-SC0005237}.

\begin{appendix}
\section{Parity Conservation Induced Joint Tunneling}
\label{sec:parity_conservation}

In this appendix, we briefly explain a striking feature caused by the presence of tMZMs: parity conservation induced \emph{joint tunneling} at both sides of the superconducting nanowire.

We consider the system near the strong coupling fixed point, where the impurity part consists of two tMZMs ($\gamma_{1,2}$) and two un-hybridized impurity Majorana fermions $(d^{\dagger}_{L,R} \!+\! d_{L,R})/\sqrt{2}$.
For the nontrivial case ($\epsilon_M \!=\! 0$), the quantum dot MZMs $(d^{\dagger}_{L,R} \!+\! d_{L,R})/\sqrt{2}$ form into singlets with their corresponding tMZMs $\gamma_{1,2}$. For simplicity, we label this state $|0\rangle_L \!\otimes\! |0\rangle_R $. We further label the exited state on each side as $|1\rangle_{L,R}$, respectively.

Now we add the weak inter-tMZM coupling $\epsilon_M$, which alters the system ground state. In this case we find the ground state of the impurity system with the Hamiltonian
\begin{equation}
H_{\text{Impurity}}' = i t_L (d^{\dagger}_L + d_L) \gamma_1 + i t_R (d^{\dagger}_R + d_R) \gamma_2 + i\epsilon_M \gamma_1 \gamma_2
\label{eq:effective_impurity_hamiltonian}
\end{equation}
through exact diagonalization. Without loss of generality, we take $t_L \!=\! t_R \!=\! t$ for simplification. The ground state has energy $\sqrt{2} t - \sqrt{\epsilon_M^2 + 8 t^2}/2$ with eigenvector
\begin{equation}
\begin{aligned}
\frac{ i \left(-2\sqrt{2} u + \sqrt{1 + 8u^2}\right) |0\rangle_L \!\otimes\! |0\rangle_R +  \, | 1\rangle_L \!\otimes\! |1\rangle_R }
{ \sqrt{1 + \left(-2 \sqrt{2} u +\sqrt{1 + 8 u^2}\right)^2}},
\end{aligned}
\label{eq:mid_impurity_es}
\end{equation}
where $u\equiv t/\epsilon_M$.
Eq.\,\eqref{eq:mid_impurity_es} correctly reduces to $|0\rangle_L \!\otimes\! |0\rangle_R$ in the limit $\epsilon_M \to 0$. 

Eq.\,(\ref{eq:mid_impurity_es}) indicates that the system ground state involves dot states with only even parity $|0\rangle_L \!\otimes\! |0\rangle_R$ and $|1\rangle_L \!\otimes\! |1\rangle_R$. Physically, this originates from the fact that the inter-tMZM coupling operator $\epsilon_M \gamma_1 \gamma_2$ changes the parity of the impurity states at \emph{both} sides.
Consequently, if we now turn on weak tunneling to the corresponding leads in both quantum dots (the resonant level model), the system allows only tunneling events that occur \emph{simultaneously} at both sides, thus doubling the scaling dimension of the leading-order terms.


\end{appendix}


\begin{thebibliography}{70}
\bibitem [{MZM()}]{MZM_pair}%
  \BibitemOpen
  \href@noop {} {}\bibinfo {note} {The 2CK quantum critical point is only well
  defined with an infinite system, where the ``partner'' of the unpaired spin
  Majorana is pushed towards infinity.}\BibitemShut {Stop}%
\bibitem [{rea()}]{real-mf}%
  \BibitemOpen
  \href@noop {} {}\bibinfo {note} {For clarification, by calling it ``real'',
  we are emphasizing that this MZM contains only fermionic impurity operators.
  This fact is in strong contrast to that of the 2CK model, whose Majorana
  fermion consists of spin operators.}\BibitemShut {Stop}%
\end{thebibliography}
\end{document}